\documentclass[aps,prd,twocolumn,nofootinbib,superscriptaddress,a4paper]{revtex4-2}
\usepackage{standalone}

\usepackage{preamble}

\usepackage[textsize=footnotesize]{todonotes}  %

\definecolor{Sorbus}{rgb}{0.996094, 0.429688, 0.0273438}

\allowdisplaybreaks

\begin{document}

\title{Nonrelativistic CFTs at Large Charge:\\ Casimir Energy and Logarithmic Enhancements}

\author{Simeon~Hellerman}
\affiliation{Kavli Institute for the Physics and Mathematics of the Universe~\textsc{(wpi)}.
The University of Tokyo
Kashiwa, Chiba 277-8582, Japan}
\author{Domenico~Orlando}
\affiliation{INFN sezione di Torino. Via Pietro Giuria 1, 10125 Torino}
\affiliation{Albert Einstein Center for Fundamental Physics, Institute for Theoretical Physics, University of Bern, Switzerland}
\author{Vito~Pellizzani}
\affiliation{Albert Einstein Center for Fundamental Physics, Institute for Theoretical Physics, University of Bern, Switzerland}
\author{Susanne~Reffert}
\affiliation{Albert Einstein Center for Fundamental Physics, Institute for Theoretical Physics, University of Bern, Switzerland}
\author{Ian~Swanson}
\noaffiliation

\date{\today}

\begin{abstract}
\noindent{}The unitary Fermi gas, by virtue of its description as a nonrelativistic conformal field theory,
has proven an interesting system by which the quantum properties of \acsp{cft} can be held to experimental verification.  Here, we examine the structure of 
conformal dimensions of charge-$Q$ operators in nonrelativistic \acsp{cft}, in the large-$Q$ 
regime, from the non-linear sigma model perspective.
We discuss in detail the renormalization of edge divergences using dimensional regularization, elucidating the presence of $\log(Q)$ terms in the large-charge expansion.
Finally, we use dimensional regularization to compute the universal one-loop $Q^0 \log Q$  contribution to the ground-state energy in $d = 3$ spatial dimensions, with
the result $\eval{\Delta(Q)}_{Q^0}  = \frac{1}{3\sqrt{3}} \log(Q) + \text{const.}$
\end{abstract}

\maketitle

\input{acronymdefinitions.acronyms}

\section{Introduction} \label{sec:introduction}

\lettrine[
    lines=8,
    lraise=0.00,
    findent=0em,
    image=true]{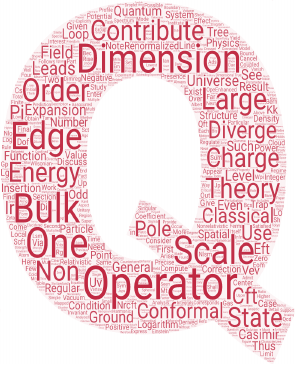}{uantum} critical points in non-relativistic systems are interesting. 
Like relativistic critical points, they are generically strongly coupled
and do not admit a uniform weak-coupling
perturbation expansion.  One
of the few analytical tools available to study them, without
modifying the theory itself, is to study the
system at large particle number, treating
the inverse particle number as a small
parameter on which to build a systematic perturbation
theory.  This approach was pursued some time ago (\emph{e.g.},~\cite{Berenstein:2002jq, Alday:2007mf, Komargodski:2012ek, Fitzpatrick:2012yx, Son:2005rv}), and presaged a 
wider exploration of the large-quantum-number regime as a
useful laboratory for studying strongly-coupled
\ac{cft}~\cite{Hellerman:2015nra, Monin:2016jmo, Cuomo:2017vzg, Sharon:2020mjs, Gaume:2020bmp, Cuomo:2020rgt, Orlando:2019hte, Cuomo:2019ejv, Orlando:2020yii, Alvarez-Gaume:2016vff, Hellerman:2017veg, Jafferis:2017zna, Alvarez-Gaume:2019biu, Kumar:2018nkf, Watanabe:2019adh, Arias-Tamargo:2019xld, Arias-Tamargo:2019kfr, Arias-Tamargo:2020fow, Badel:2019oxl, Badel:2019khk, Giombi:2020enj, Antipin:2021akb, Cuomo:2021qws, Komargodski:2021zzy, Cuomo:2021ygt, Favrod:2018xov, Kravec:2018qnu, Kravec:2019djc, Orlando:2020idm, Hellerman:2020eff, Pellizzani:2021hzx, Jack:2021ypd, Dupuis:2021yej, Dondi:2021buw, Hellerman:2017sur, Bourget:2018obm, Hellerman:2018xpi, Beccaria:2018xxl, Beccaria:2018owt, Beccaria:2020azj, Grassi:2019txd,  Antipin:2020abu, Antipin:2020rdw, Hellerman:2020sqj, Hellerman:2021yqz, Hellerman:2021duh, Fiol:2021icm, Antipin:2021jiw, Aharony:2021mpc, Dupuis:2021flq, Cuomo:2021cnb, Jack:2021lja, Antipin:2021rsh, Orlando:2021usz, Giombi:2021zfb}.

One exciting application emerged with the study of the 
so-called %
unitary Fermi gas, wherein interacting fermions
at zero temperature are described by a \ac{nrcft}
when the $s$-wave scattering length becomes infinite \cite{Nishida:2010tm}.
This is realized
in the laboratory by trapping a fermion species in a confining potential and tuning a background magnetic field to a
Feshbach resonance~\cite{Giorgini_2008}. The energy levels of such a system are mapped bijectively to the nonnegatively-charged operator dimensions of the theory without the trap, through the nonrelativistic state-operator correspondence~\cite{Werner_2006,Nishida:2007pj,Goldberger:2014hca}. Thus, by extracting the universal properties of operator dimensions in sectors of fixed charge $Q\gg1$, we obtain predictions that yield to direct experimental verification.

It was only recently, however, 
that the large-charge expansion of operator dimensions was pursued
systematically at subleading orders, in part
because of the technical complications
involved with short-distance
singularities at the edge
of the particle droplet (see~\cite{Son:2005rv,Kravec:2018qnu,Kravec:2019djc,Orlando:2020idm} for preliminary work on the subject). The development
of a Wilsonian description of operators localized at the edge of the
droplet~\cite{Hellerman:2020eff}
(utilizing the general analysis of \ac{cft} boundary operators
at large quantum number~\cite{Hellerman:2016hnf,Cuomo:2021cnb})
has made possible a systematic analysis of
the renormalization of these divergences
in large-charge perturbation theory. 
This procedure was generalized to arbitrary spatial dimension in~\cite{Pellizzani:2021hzx}, revealing a subtle structure involving $\log(Q)$ terms in the large-$Q$ expansion of the lowest operator dimension. 
It is useful to highlight a few outcomes of this program in some specific, terrestrially relevant examples.
In $d=2$, the spectrum was described in~\cite{Hellerman:2020eff}, up to and including order $Q^0$, in terms
of four unknown Wilsonian coefficients of bulk and edge operators. In particular, it was shown that the $Q^0$ term is theory-independent, 
and can be computed numerically, to arbitrary precision, 
with the result \(-0.294159\dots\), which was independently confirmed by $\zeta$-function regularization in~\cite{Orlando:2020idm}.
In odd spatial dimensions, and in particular in $d=3$, it was shown that there are bulk operators in the large-charge
\ac{eft} arising at order $Q^0$, so the order-$Q^0$ term is not universal~\cite{Pellizzani:2021hzx}.
This is precisely analogous to the case of relativistic \ac{cft} at
large charge~\cite{Orlando:2019hte, Orlando:2019skh,Orlando:2020yii} in 
four spacetime dimensions. However,
it was pointed out in~\cite{Cuomo:2020rgt} that this statement belies a crucial
structural aspect of the theory: The nonuniversality 
of the $Q^0$ term \emph{is related directly to the presence of a
$Q^0 \log(Q)$ term in the expansion, with a universal coefficient}.  This holds in any even spacetime dimension.  In this paper, we show explicitly that this applies also to the large-$Q$ expansion of \ac{nrcft}.  To achieve this, we are charged with the task
of elucidating the classical and quantum logarithmic enhancements to the spectrum emerging
from the aforementioned structure.  Crucially, we compute the universal coefficient of the $Q^0 \log(Q)$ term in $d=3$ spatial dimensions, which characterizes the Casimir contribution to 
the energy spectrum above the ground state. 

The organization of the paper is as follows.
In Section~\ref{sec:effAction} we discuss the general structure of the large-charge expansion of \acp{nrcft}, distinguishing the contributions to the conformal dimensions due to the  insertions of bulk and edge operators.
In Section~\ref{sec:logarithmic-terms} we discuss the logarithmic terms in the expansion, classified by tree-level operator insertions (\S~\ref{sec:logs-from-tree}) and one-loop quantum terms (\S~\ref{sec:universal-one-loop}).
In Section~\ref{sec:multi-vertex} we discuss the connected multi-vertex contributions to the classical solution, showing that such corrections only enter at negative order in~\(\mu\), and thus do not affect the results of our analysis of terms with nonnegative powers of $\mu$.

\section{Large charge effective action}
\label{sec:effAction}
\subsection{Heuristics}

\noindent{}The theories of interest are invariant under the nonrelativistic incarnation of the conformal group, the \emph{Schrödinger} group, which includes a $U(1)$ central extension that corresponds to particle-number conservation. Such theories 
admit a tractable semi-classical description when the associated particle number $Q$ is large~\cite{Son:2005rv,Favrod:2018xov,Kravec:2018qnu,Kravec:2019djc,Orlando:2020idm,Hellerman:2020eff,Pellizzani:2021hzx}.
This regime is achieved by taking advantage of the nonrelativistic state-operator correspondence, whereby the (charged) operator spectrum is mapped bijectively to the energy spectrum of the system subjected to an external harmonic potential $A_0(\vec x) = \frac{m \omega^2}{2 \hbar} |\vec x|^2$~\cite{Werner_2006,Nishida:2007pj,Goldberger:2014hca}, with trapping frequency $\omega$. Explicitly, with $\hbar = m = 1$, we have
\begin{equation}
  E_0 = \omega \cdot \Delta(Q) \ ,
\end{equation}
where $\Delta(Q)$ is the lowest operator dimension of charge $Q$, and $E_0$ is the ground-state energy of the system with $Q$ particles in the trap.  Borrowing intuition from quantum mechanical systems in confining potentials and turning points, it is clear that the particle density decreases as we recede from the center of the trap.  The resulting droplet of trapped particles occupies a finite region of space that is characterized by the strength $\omega$ of the potential and the charge $Q$ itself. As we shall see, the droplet is supported over a region that is classically spherical, with radius $R_{\text{cl}} \sim Q^{1/(2d)}/\sqrt{\omega}$.
Note that this structure is distinct from the \emph{relativistic} state-operator correspondence, where the theory is mapped to a rigid cylinder, and the superfluid configuration in the ground state is homogeneous~\cite{Hellerman:2015nra,Monin:2016jmo}. One important convenience in this construction is that trapping potentials used in ultracold atom experiments are well approximated by a harmonic potential~\cite{Dalfovo_1999,Giorgini_2008}, so that $E_0$ is an experimentally accessible observable.  Moreover, the Thomas-Fermi approximation commonly used to fit these data holds when the number of particles is large, corresponding to the leading-order, large-charge prediction.
To get a feeling for how things scale, consider the charge and energy densities 
in the bulk (%
\emph{i.e.}, sufficiently far from the edge of the cloud):
\begin{equation}
\begin{aligned}
	\text{charge density} & \sim \frac{Q}{\mathrm{Vol}} \sim \omega^{d/2} \sqrt{Q}\ , \\
	\text{energy density} & \sim \frac{E_0}{\mathrm{Vol}} \sim {\omega^{d/2+1}} \Delta(Q)/{\sqrt{Q}}\ ,
\end{aligned}
\end{equation}
where the volume of the cloud scales as $\mathrm{Vol} \sim R_{\text{cl}}^d \sim \sqrt{Q}/\omega^{d/2}$.  We can take the limit  $Q \to \infty$ and $\omega\to 0$ with the charge density held fixed.  In this limit we expect the energy density will remain fixed as well, and we should recover a homogeneous ground state.  For this limit to exist, the scaling of $\Delta(Q)$ has to go like
\begin{equation} \label{eq:DeltaQ_LOsimple}
    \Delta(Q) \sim Q^{(d+1)/d}\ , 
\end{equation}
to leading order.
The low-energy effective theory for the Goldstone mode $\chi$ associated with the spontaneous breaking of the $U(1)$ symmetry is characterized by the \ac{ir} scale $R_{\text{cl}}^{-1}$ and the \ac{uv} scale $\rho^{1/d}$, where $\rho$ is the charge density above.
The derivative expansion in the bulk \ac{eft} is thus controlled by
\begin{equation}
    {R_{\text{cl}}^{-1}}/{\rho^{1/d}} \sim Q^{-1/d}\ .
\end{equation}
This ratio needs to be small for the effective description to be tractable, which requires the charge to be large. Assuming parity invariance, corrections to Eq.~(\ref{eq:DeltaQ_LOsimple}) are in fact controlled by the square of this ratio.
While this seems to reproduce exactly the relativistic result~\cite{Hellerman:2015nra}, an entirely new structure arises from the boundary dynamics of the system~\cite{Hellerman:2020eff,Pellizzani:2021hzx}. The edge of the cloud is characterized by the vanishing of the particle density $\rho$, which can be thought of as a Dirichlet boundary condition, where the location of the boundary can itself fluctuate. The breakdown of the bulk \ac{eft} in this region gives rise to divergences in the theory, which can be controlled by the presence of operators localized at the edge~\cite{Hellerman:2020eff}. The first edge contribution to $\Delta(Q)$ enters at order $Q^{(2d-1)/(3d)}$, and the role of the dressing field/\ac{uv} scale is played by
the \emph{gradient} of $\rho^{1/d}$ rather than $\rho^{1/d}$ itself. Consequently, the expansion parameter scales like $Q^{-2/(3d)}$.
Moreover, the cancellation of edge divergences is resp
onsible for the presence of $\log(Q)$ terms in the expansion of $\Delta(Q)$. 
This yields the general structure~\cite{Pellizzani:2021hzx}
\begin{equation}
\begin{aligned}
	\Delta(Q) ={}& Q^{1 + 1/d} \left[ a_1 + \frac{a_2}{Q^{2/d}} + \frac{a_3}{Q^{4/d}} + \dots \right] \\
	& + Q^{2/3 - 1/(3d)} \left[ b_1 + \frac{b_2}{Q^{2/(3d)}} + \frac{b_3}{Q^{4/(3d)}} + \dots \right] \\
           & + Q^{1/3 - 5/(3d)} \left[ d_1 + \frac{d_2}{Q^{2/(3d)}} + \frac{d_3}{Q^{4/(3d)}} + \dots \right] \\
	& + \text{\(\log(Q)\) enhancements.}
\end{aligned}
\end{equation}
In the next section, we discuss this in detail and recover these results of from the \ac{nlsm} perspective using \ac{dimreg} and we clarify when the $\log(Q)$ terms occur.

\subsection{Leading order}

\noindent{}Consider the leading-order Lagrangian\footnote{It is convenient to let $\omega = 1$ (along with $\hbar = m = 1$).} of the \ac{nlsm} in $d$ spatial dimensions~\cite{Son:2005rv,Favrod:2018xov,Kravec:2018qnu}:
\begin{equation}%
\label{eq:LO-action}
	\mathcal{L}_{LO} = c_0 U^{d/2+1}\ ,
\end{equation}
where
\begin{equation}
	U = \dot \chi - \frac{1}{2} r^2 - \frac{1}{2} (\del_i \chi)^2\ .
\end{equation}
The harmonic trap $A_0(r) \equiv \frac{1}{2} r^2$ (with $r = \abs{\vec x}$) is coupled to $\dot\chi$, respecting general coordinate invariance~\cite{Son:2005rv}.
Here, $\chi$ is the only massless low-energy degree of freedom present in the theory, \emph{i.e.}, the Goldstone boson associated with the spontaneous breaking of $U(1)$. In the superfluid ground state, it acquires the usual \ac{vev} $\ev{\chi} = \mu \cdot t$, where $\mu$ is referred to as the \emph{chemical potential}. For comparison, the equivalent Lagrangian in the relativistic case would read $\mathcal{L} = c_0 (\del \chi)^{d+1}$~\cite{Hellerman:2015nra,Monin:2016jmo}.

In the ground state, $U$ takes the \ac{vev}
\begin{equation}
	\ev{ U } = \mu - \frac{1}{2} r^2 \equiv \mu \cdot z\ ,
\end{equation}
where \(z\) is a dimensionless coordinate $z \equiv 1 - r^2/(2\mu) \equiv 1 - r^2/R_{\text{cl}}^2$, with the classical radial boundary $R_{\text{cl}} \equiv \sqrt{2 \mu}$. The ground-state charge density is
\begin{equation}
	\ev{ \rho }  = \ev{ \pdv{\mathcal{L}_{LO}}{\dot \chi} } = \ev{\pdv{\mathcal{L}_{LO}}{ U}} \sim  \ev{U^{d/2}} \sim (\mu \cdot z)^{d/2}\ .
\end{equation}
Classically, the ground-state density is thus seen to decrease away from the origin ($z=1$), and vanishes when $z=0$, \emph{i.e.,} $r = R_{\text{cl}}$, which indicates that the particles are confined in a cloud (or droplet) that is classically spherically symmetric with radius $R_{\text{cl}}$. The total charge is obtained by integrating the charge density over the volume of the cloud, relating $Q$ to the chemical potential:
\begin{equation} \label{eq:mu_LO}
	\mu =  \frac{1}{\sqrt{2 \pi}} \left[ \frac{\Gamma(d+1)}{\Gamma\left( \frac{d}{2} + 2 \right) c_0} \right]^{1/d} Q^{1/d} \equiv \zeta Q^{1/d} \ .
\end{equation}
It follows that the ground-state energy --- and, therefore, the dimension of the lowest operator of charge $Q$ --- is
\begin{equation} 
\label{eq:DeltaQ_LO}
	\Delta(Q) = \frac{d}{d+1} \zeta Q^{(d+1)/d}\ .
\end{equation}

Since the ground state preserves spherical symmetry, it is convenient to express every \ac{vev} in terms of $z$, which simplifies the computation of Eqns.~(\ref{eq:mu_LO}) and (\ref{eq:DeltaQ_LO}).  In fact, we shall always do so, using the following properties:
\begin{equation} \label{eq:z_stuff}
\begin{aligned}
	 (\del_i f(\abs{\vec x})) (\del_i g(\abs{\vec x})) &= \frac{2(1-z)}{\mu} f'(z) g'(z)\ , \\
	 \nabla^2 f(\abs{\vec x}) &= \frac{2}{\mu} \left[ (1-z) f''(z) - \frac{d}{2} f'(z) \right]\ , \\
	 \int_{\text{cloud}} \dd[d]{x} \, f(\abs{\vec x}) &= \frac{(2 \pi \mu)^{d/2}}{\Gamma\left(\frac{d}{2}\right)} \int_0^1 \dd{z} (1 - z)^{d/2-1} f(z)\ ,
\end{aligned}
\end{equation}
where primes refer to derivatives with respect to $z$, and both $f$ and $g$ are spherically invariant.

\subsection{Higher-order terms}
\label{sec:higher-order-terms}

\noindent{}Besides $U$ and its derivatives, the only operator allowed by general coordinate invariance with a nonvanishing \ac{vev} in the superfluid ground state is
\begin{equation}
	Z = \nabla^2 A_0 - \frac{1}{d} (\nabla^2 \chi)^2\ ,
\end{equation}
with $\ev{ Z } = d$. Therefore, in the bulk, all nontrivial operators are composite operators made out of integer powers of $(\del_i U)^2$ and $Z$, which are then dressed to marginality with an appropriate (possibly fractional) power of $U$~\cite{Pellizzani:2021hzx}:
\begin{equation}
	\Op{m}{n} \equiv c_{m,n} \cdot (\del_i U)^{2m} Z^n U^{d/2+1-(3m+2n)}\ ,
\end{equation}
where $m$ and $n$ are integers, and we have explicitly incorporated the Wilsonian coefficients $c_{m,n}$. This structure can also be recovered using the coset construction~\cite{Kravec:2018qnu}, wherein the aforementioned dressing rule comes from integrating out (via an inverse Higgs constraint) the massive dilaton mode.

Using Eq.~(\ref{eq:z_stuff}), we see that the contributions to the charge and to the conformal dimension from the insertion of \(\Op{m}{n}\) are respectively given by%
\begin{widetext}
\begin{equation} \label{eq:QMN}
	Q^{(m,n)}  \equiv \int_{\textnormal{cloud}} \dd^d x \, \left\langle \frac{\partial \Op{m}{n}}{\partial U} \right\rangle 
   =  \frac{2^m d^n (2 \pi)^\frac{d}{2}}{\Gamma\left(\frac{d}{2}\right)} \mu^{d-2(m+n)}  \frac{\Gamma\left( \frac{d}{2} + m \right) \Gamma\left( \frac{d}{2} + 2 - (3 m + 2 n) \right)}{\Gamma(d + 1 - 2 (m + n))},
\end{equation}
and
\begin{equation} \label{eq:DeltaMN}
\begin{aligned}
	\Delta^{(m,n)}  &= \int_{\textnormal{cloud}} \dd^d x \, \left\langle \frac{\partial \Op{m}{n}}{\partial U} \dot\chi - \Op{m}{n} \right\rangle \\
	& = \frac{2^m d^n (2\pi)^\frac{d}{2}}{\Gamma\left(\frac{d}{2}\right)} \mu^{d+1-2(m+n)} \cdot \frac{\Gamma\left( \frac{d}{2} + m \right) \Gamma\left( \frac{d}{2} + 2 - (3 m + 2 n) \right)}{\Gamma(d + 1 - 2 (m + n))} \cdot \frac{d - 2 (m + n)}{d + 1 - 2 (m + n)}.
\end{aligned}
\end{equation}
\end{widetext}
where we have used the analytic continuation of the Beta function. Note that the $(0, 0)$-contribution gives back Eqs.~\eqref{eq:mu_LO} and \eqref{eq:DeltaQ_LO}.
Since the large-charge regime corresponds to the large-\(\mu\) regime\footnote{Had we kept track of the strength $\omega$ of the potential, we would have had $\mu\gg\omega$.},
it is convenient to identify the leading \(\mu\)-scaling of the integrated operator in the conformal dimension:
\begin{equation}
  \label{eq:mu-scaling-O-mn}
  \mudim{\Op{m}{n}} \equiv d+1-2(m+n)\ ,
\end{equation}
where we have introduced the notation $\mudim{\cdot}$ to extract the scaling exponent.

Together with these bulk operators, one must consider insertions localized at the edge of the cloud~\cite{Hellerman:2020eff}, which, as we will discuss in detail in the following, are needed to regulate possible divergences in Eq.~\eqref{eq:DeltaMN}.
The general form of an edge operator is~\cite{Pellizzani:2021hzx}
\begin{equation} \label{eq:Zp_edge}
  \Z{p} \equiv \kappa_p  Z^p  \delta(U)  (\del_i U)^{(d+4(1-p))/3}\ ,
\end{equation}
where $p$ is an integer, $\kappa_p$ is a Wilsonian coefficient, and $\delta(U)$ is an operator-valued $\delta$-function associated with the vanishing of the particle density. Such an operator contributes to the total charge and the conformal dimension, respectively, as follows:
\begin{equation} \label{eq:QP}
\begin{aligned}
	Q^{(p)} & \equiv \int_{\textnormal{cloud}} \dd^d x  \left\langle \frac{\partial \Z{p}}{\partial U} \right\rangle \\
	 & = \frac{2^\frac{d+4(1-p)}{6} d^p (2 \pi)^\frac{d}{2}}{\Gamma\left(\frac{d}{2}\right)} \mu^\frac{2 d - 4 - 2 p}{3}  \int_0^1 \dd z \, \delta'(z) (1 - z)^\frac{2d-1-2p}{3} \\
	& = \frac{2^\frac{d+4(1-p)}{6} d^p (2 \pi)^\frac{d}{2}}{\Gamma\left(\frac{d}{2}\right)} \mu^\frac{2 d - 4 - 2 p}{3}  \frac{2 d - 1 - 2 p}{3},
\end{aligned}
\end{equation}
and
\begin{equation} \label{eq:DeltaP}
\begin{aligned}
	\Delta^{(p)} & \equiv \int_{\textnormal{cloud}} \dd^d x \, \left\langle \frac{\partial \Z{p}}{\partial U} \dot\chi - \Z{p} \right\rangle \\
	& = \frac{2^\frac{d+4(1-p)}{6} d^p (2 \pi)^\frac{d}{2}}{\Gamma\left(\frac{d}{2}\right)} \mu^\frac{2d-1-2p}{3} \cdot \frac{2 d - 4 - 2 p}{3},
\end{aligned}
\end{equation}
where a prime refers to a derivative with respect to $z$, and $\delta'(z)$ is defined via integration by parts. As above, we denote the leading \(\mu\)-scaling of the edge operator as
\begin{equation} \label{eq:Zp_muscaling}
  \mudim{\Z{p}} = \frac{2(d -p)-1}{3}\ .
\end{equation}

Summing all the possible contributions to the conformal dimension and the charge and eliminating \(\mu\), we find that the expression of \(\Delta = \Delta(Q)\) takes the form of a Puiseux series~\cite{siegel1969topics}
\begin{equation}
\begin{aligned}
	\Delta(Q) ={}& Q^{1 + 1/d} \left[ a_1 + \frac{a_2}{Q^{2/d}} + \frac{a_3}{Q^{4/d}} + \dots \right] \\
	& + Q^{2/3 - 1/(3d)} \left[ b_1 + \frac{b_2}{Q^{2/(3d)}} + \frac{b_3}{Q^{4/(3d)}} + \dots \right] \\
           & + Q^{1/3 - 5/(3d)} \left[ d_1 + \frac{d_2}{Q^{2/(3d)}} + \frac{d_3}{Q^{4/(3d)}} + \dots \right] \\
	& + \text{\(\log(Q)\) enhancements,}
\end{aligned}
\end{equation}
where the third line is due to the non-linear interactions between the contributions of the bulk and edge operators.
The appearence of logarithmic terms in the conformal dimension is related to the renormalization of edge divergences using edge operators of the form in Eq.~(\ref{eq:Zp_edge}).
Such a divergence was first observed in~\cite{Kravec:2018qnu} for \(d=2\), and subsequently renormalized in~\cite{Hellerman:2020eff} to find
\begin{multline}
  \eval{\Delta(Q)}_{d=2} = \frac{2 Q^\frac{3}{2}}{3 \sqrt{2 \pi c_0}} - \sqrt{\frac{8 \pi}{c_0}} Q^\frac{1}{2} (c_1 \log Q + \kappa_0^\text{ren}) \\
  - \left( \frac{2 \pi^5}{c_0} \right)^\frac{1}{6} 4 \kappa_1 Q^\frac{1}{6} - 0.29416 + \mathcal{O}\left( Q^{-\frac{1}{6}} \right)   
\end{multline}
In what follows, we will generalize this observation.

\section{Logarithmic terms}
\label{sec:logarithmic-terms}

\subsection{Logarithms from tree-level insertions}
\label{sec:logs-from-tree}

\noindent{}We have seen in Eq.~\eqref{eq:DeltaMN} that an insertion of \(\Op{m}{n}\) contributes to the conformal dimension to leading-order as
\begin{multline} \label{eq:DeltaMN_bis}
    \Delta^{(m,n)} = C_{m,n}^d \mu^{d+1-2(m+n)}  \\
     \times  \frac{\Gamma\left( \frac{d}{2} + m \right) \Gamma\left( \frac{d}{2} + 2 - (3 m + 2 n) \right)}{\Gamma\left( d + 2 - 2 (m + n) \right)}\ ,
\end{multline}
where $C_{m,n}^d$ is a $\mu$-independent constant that incorporates the Wilsonian coefficient and numerical factors coming from the integration over the volume. Similarly, the leading-order contribution of an edge operator $\Z{p}$ is
\begin{equation}
    \Delta^{(p)} = K_p  \mu^{(2(d-p)-1)/3}\ ,
\end{equation}
where $K_p$ is a factor similar to $C_{m,n}^d$ above.

The gamma function has no zeros, but has single poles when the argument is a non-positive integer:
\begin{multline}
   \Gamma(-k + \epsilon) = \frac{(-1)^k}{k!} \pqty{\frac{1}{\epsilon} + H_k - \gamma} + \order{\epsilon}\ ,\\
   k = 0,1, \dots,
\end{multline}
where \(H_k = \sum_{l=1}^n \tfrac{1}{l}\) is the \(k\)-th harmonic number.
This means that there are values of \(m\) and \(n\) such that the contribution of \(\Op{m}{n}\) is singular and needs to be renormalized by the insertion of an edge operator.
These are the \emph{edge divergences} of~\cite{Son:2005rv,Kravec:2018qnu,Orlando:2020idm}, and a physically motivated renormalization procedure was discussed in~\cite{Pellizzani:2021hzx}, based on~\cite{Hellerman:2020eff}.
However, the analytic continuation in terms of gamma functions captures the same physical content while dramatically simplifying the computations, leaving us with the simpler task of identifying the locations of the poles and demonstrating that there always exists an appropriate counterterm for each pole.

\subsubsection{Location of the poles}

\begin{figure}
  \centering
  \includestandalone[mode=buildnew]{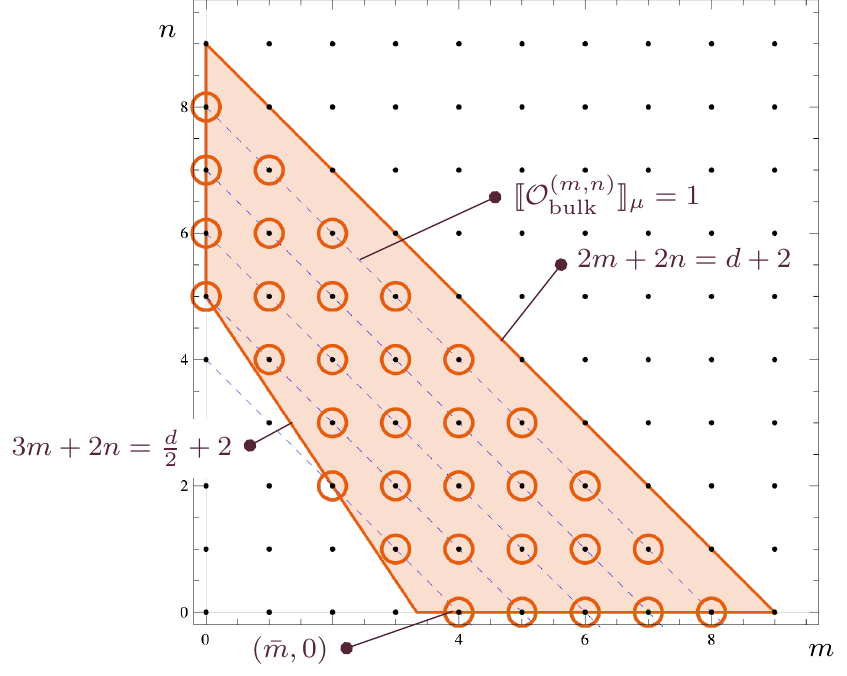}
  \caption{Divergent bulk operators \(\Op{m}{n}\). The operators corresponding to lattice points between the lines \( 2 m + 2 n = d + 2 \) and \(3m + 2n =\frac{d}{2} + 2\) have a simple pole and need to be regulated (shaded region).
    All the operators on a line \(m + n  = \text{const.}\) have the same \(Q\)-scaling (dashed lines).
  The operator \(\Op{\bar m}{0}\) has the highest \(Q\)-scaling.}
  \label{fig:Lines-and-dots}
\end{figure}

\begin{table}
  \centering
  \begin{tabular}{p{1em}lcc}
    \toprule
    \(d\) & \((m,n)\) & \(\mudim{\Op{m}{n}}\)  \\
    \midrule
    \(2\) & \((1,0)\) & \(1\) & \ \raisebox{-.6\height}{\includegraphics[width=3em]{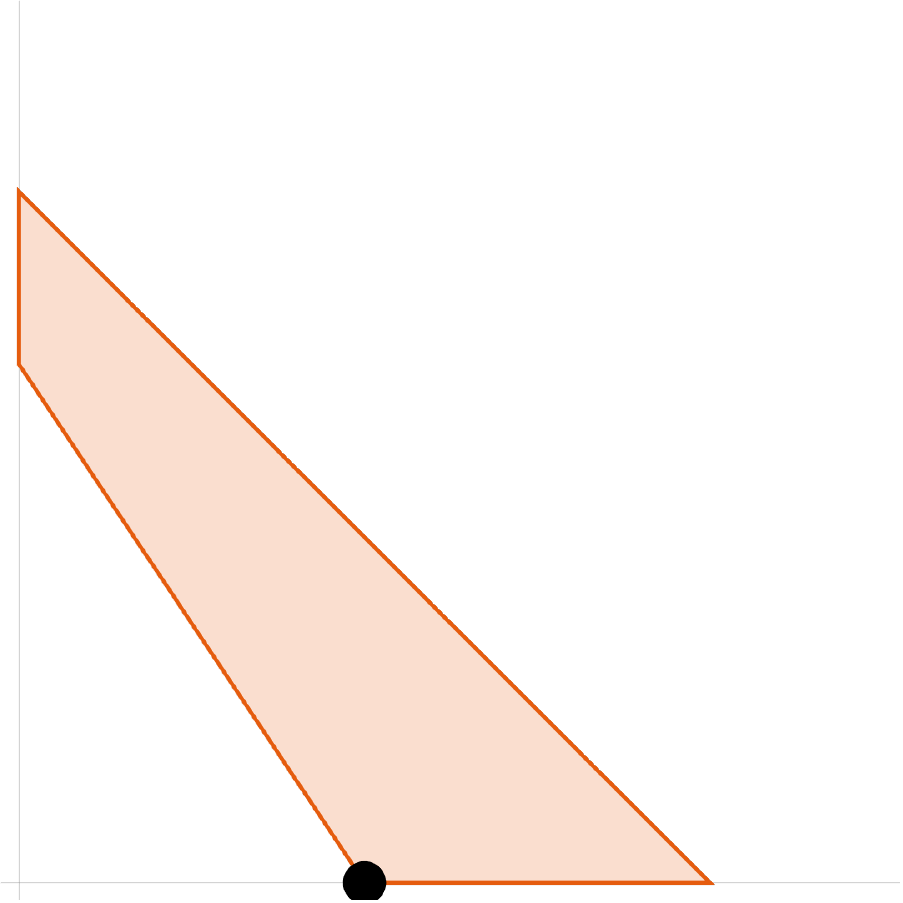} } \\[1em]
    \(4\) & \((2,0)\), \((1,1)\), \((0,2)\) & \(1\) & \raisebox{-.6\height}{\includegraphics[width=3em]{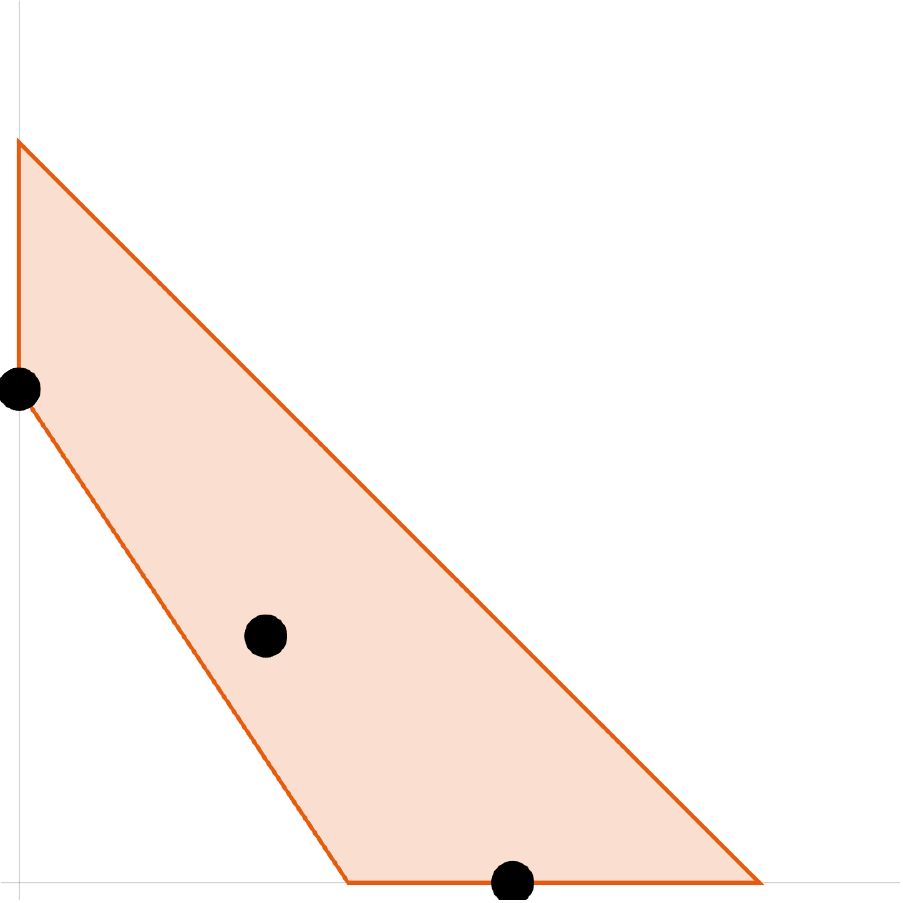}} \\[1em]
    \(6\) & \((2,0)\), \((1,1)\) & \(3\) & \multirow{2}{*}{\includegraphics[width=3em]{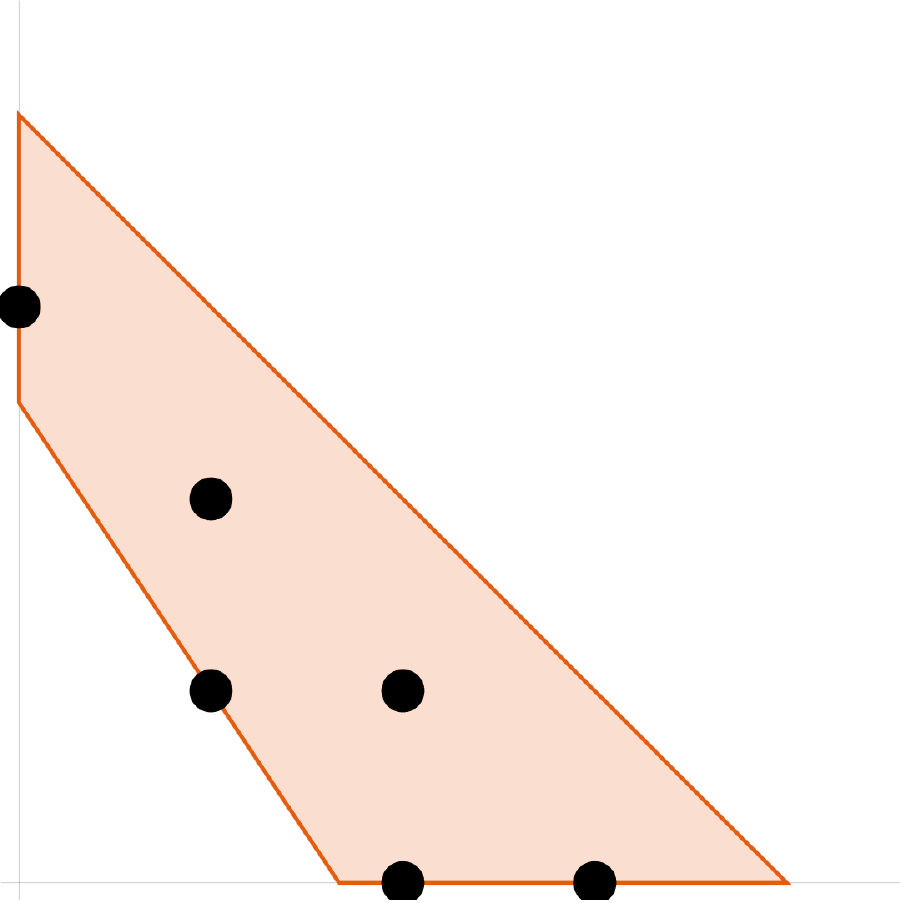}} \\ 
          & \((0,3)\), \((1,2)\), \((2,1)\), \((3,0)\) & \(1\) \\[1em]
    \(8\) & \((2,0)\) & \(5\) & \multirow{3}{*}{\includegraphics[width=3em]{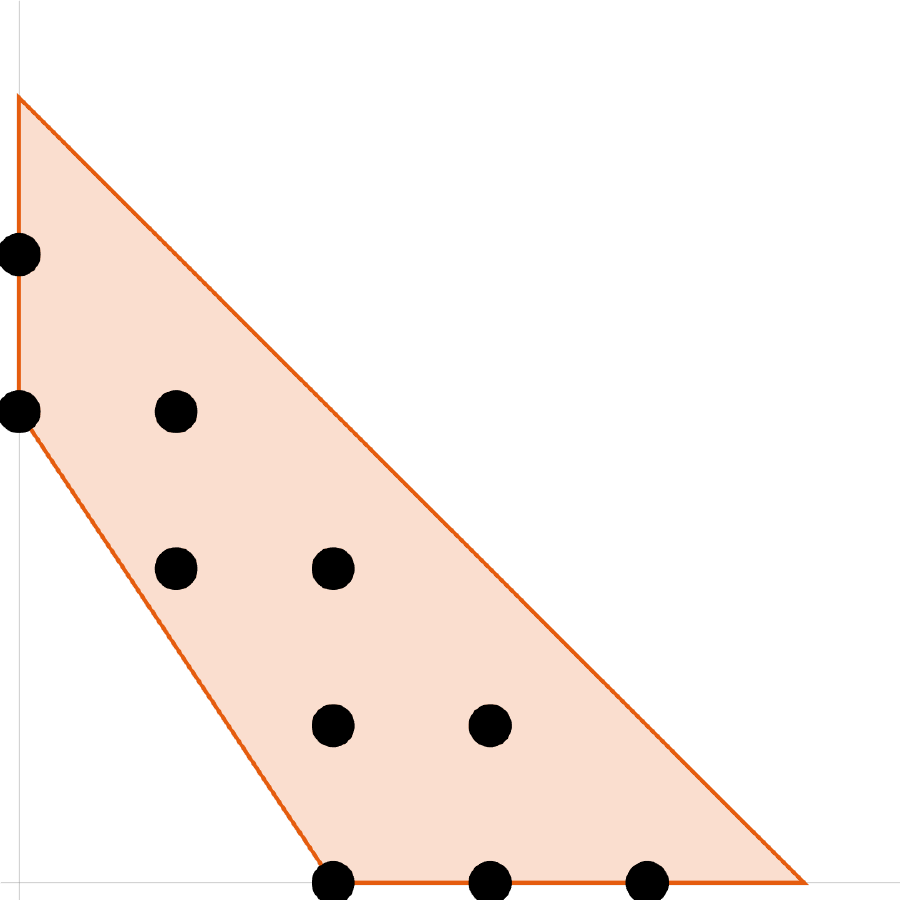}} \\
          & \((0,3)\), \((1,2)\), \((2,1)\), \((3,0)\) & \(3\) \\
          & \((0,4)\), \((1,3)\), \((2,2)\), \((3,1)\), \((4,0)\) & \(1\)\\
    \bottomrule
  \end{tabular}
  \caption{Divergent bulk operators  in different dimensions.}
  \label{tab:divergent-bulk-operators}
\end{table}

\noindent{}Let us enumerate which values of \(m\) and \(n\) give rise to poles, and at what order in \(\mu\) they enter the conformal dimension. 
From the expression of \(\Delta^{(m,n)}\) in Eq.~\eqref{eq:DeltaMN_bis}, we see that there is a pole every time the argument of the second gamma function in the numerator is a non-positive integer, and the argument of the gamma function in the denominator is a positive integer (\emph{i.e.,} every time the numerator exhibits a pole that is not canceled by a pole in the denominator):
\begin{equation}
  \label{eq:bulk-divergence-conditions}
  \begin{cases}
    \frac{d}{2} + 2 - (3 m + 2 n) = 0, -1, -2, \dots \\
     d + 2 - 2 (m + n) = 1, 2, 3, \dots
  \end{cases}
\end{equation}
The solutions to these equations for the first few values of \(d\) are given in Table~\ref{tab:divergent-bulk-operators} (see also Figure~\ref{fig:Lines-and-dots}).
The first condition can be satisfied only if \(d\) is even: \emph{There are no poles from tree-level insertions in odd spatial dimension}.\footnote{It is not clear to us if there is a simple or intuitive interpretation of this general rule.}
The second condition can be rewritten using the \(\mu\)-scaling of the operator \(\Op{m}{n}\) as \(d+2-2(m+n) = \mudim{\Op{m}{n}} + 1 = 1, 2, 3 \dots\)
However, since \(d \) must be even, the $\mu$-scaling must be odd and positive:
\begin{equation} \label{eq:evenD_logScaling}
  \mudim{\Op{m}{n}} = d + 1 - 2 (m + n ) = 1, 3, 5, \dots
\end{equation}
We thus recover another general property: \emph{Only operators of positive \(Q\)-scaling can diverge at tree level}. %

The largest possible \(\mu\)-scaling corresponds to the operator \(\Op{\bar m}{0}\), where \((\bar m,0)\) is the lattice point closest to the line \(d/2 + 2 - 3 m - 2n= 0\) (see Figure~\ref{fig:Lines-and-dots}):
\begin{equation}
  \bar m = \left\lceil \frac{d}{6} + \frac{2}{3}   \right\rceil\ ,
\end{equation}
so that
\begin{equation}
  \mudim{\Op{m}{n}} \le \mudim{\Op{\bar m}{0}} = d + 1 - 2  \left\lceil \frac{d}{6} + \frac{2}{3}   \right\rceil\ .
\end{equation}
Note, however, that in general there will be different operators that contribute at the same order in \(\mu\) to the conformal dimension.
In fact, while the highest scaling grows like \(\mudim{\Op{\bar m}{0}} \sim 2d/3 \), the number of diverging operators grows quadratically with \(d\),
\begin{multline}  
  N_{\rm div}(d) = \frac{5 d^2}{48} + \frac{5 d}{12} + \frac{1}{3} \abs{\left( d-3 \left\lfloor \frac{d}{3}\right\rfloor -1\right)} \\
   + \frac{(-1)^{d/2}}{8} -\frac{11}{24} \ .
\end{multline}
The important question is now whether there exists an edge operator \(\Z{p}\) for each divergent bulk insertion.

\subsubsection{Counterterms and renormalization}

\noindent{}Whenever one or more bulk operators \(\Op{m}{n}\) give a singular contribution to the operator dimension $\Delta(Q)$, we need to regulate them via the insertion of an edge counterterm \(\Z{p}\) of the same \(\mu\)-scaling~\cite{Hellerman:2020eff,Pellizzani:2021hzx},
\begin{equation}
\label{eq:edge-regulation}
    \mudim{\Z{p}} = \mudim{\Op{m}{n}} \ .
\end{equation}
If such a term exists (\emph{i.e.,} if the condition is satisfied for a non-negative integer \(p\)), its bare coupling $\kappa_p$ is renormalized to cancel the divergence. More explicitly, by Eq.~(\ref{eq:mu-scaling-O-mn}) and (\ref{eq:Zp_muscaling}), we need to impose
\begin{equation}
\label{prelimConditionForBulkAndEdgeMapping}
  p %
  = n - \left[\frac{d}{2} + 2 - \left(3m+2n\right) \right] \equiv n - k\ .
\end{equation}
We have to check whether this condition can be satisfied whenever $\Op{m}{n}$ induces a pole (and thus a logarithm), i.e. whenever $d$, $m$ and $n$ satisfy Eq.~\eqref{eq:bulk-divergence-conditions}. If so, then every bulk divergence can be cured by an appropriate counterterm \(\Z{p}\).

In fact, it is easy to see that the condition for a pole to appear is strictly \emph{stronger} than the condition for an edge counterterm to have the appropriate $\mu$-scaling to renormalize it, namely Eq.~\eqref{prelimConditionForBulkAndEdgeMapping} above. Indeed, the first condition in Eq.~\eqref{eq:bulk-divergence-conditions} is enough to guarantee that Eq.~\eqref{prelimConditionForBulkAndEdgeMapping} is satisfied for some non-negative integer $p$, since $n$ is a positive integer and $k$ is a non-positive integer in this case. Also, recall that this only happens in even $d$. As already mentioned in Eq.~\eqref{eq:evenD_logScaling}, the second condition for a pole to appear simply translates into $\mudim{\Op{m}{n}} = 1, 3, 5, \ldots$. In short, since the condition 
for the presence of poles is more restrictive than Eq.~(\ref{prelimConditionForBulkAndEdgeMapping}), there 
will always exist edge terms with the same $\mu$-scaling for all (tree level) log-inducing bulk insertions, and the latter necessarily exhibit positive $\mu$-scaling $\mudim{\Op{m}{n}} \in \setN$.

\bigskip

The precise mechanism of the cancellation of divergences between bulk and edge operators can be easily understood in \ac{dimreg}. In \(d + 2 \epsilon\) dimensions, let us consider all divergent bulk operators \(\Op{m}{n}\) with the same $\mu$-scaling, and let us denote their cumulative leading-order contribution to the conformal dimension by
\begin{equation}
  \Delta^{(m,n)} = \mu^{d+1 - 2(m+n) + 2 \epsilon} \cdot C_{m,n}^{d+2\epsilon}\, \Gamma(-k + \epsilon)\ , 
\end{equation}
where the coefficient \(C_{m,n}^{d+2\epsilon}\) is regular for \(\epsilon \to 0\).
For small values of \(\epsilon\), we have
\begin{widetext}
  \begin{equation} 
      \Delta^{(m,n)} = \mu^{d+1 - 2(m+n) + 2 \epsilon} \cdot \frac{(-)^k C_{m,n}^{d+2\epsilon}}{k!} \pqty{\frac{1}{\epsilon} + H_k -\gamma + \order{\epsilon}}
      = \mu^{d+1 -2 (m+n)} \frac{(-)^k C_{m,n}^d}{k!} \pqty{\frac{1}{\epsilon} +  2 \log(\mu) + \text{cst.}} + \order{\epsilon} \ .
  \end{equation}
\end{widetext}
The pole at \(\epsilon \to 0\) is eliminated by renormalizing the bare coupling $\kappa_p$ of the edge operator \(\Z{p}\) satisfying Eq.~(\ref{eq:edge-regulation}) as $\kappa_p = \kappa_p^{\text{ren}} - \kappa_p^{(-1)}/\epsilon $, where $\kappa_p^{\text{ren}}$ is regular and \(\kappa_p^{(-1)}\) is chosen so that the corresponding contribution to the conformal dimension \(\Delta^{(p)}\) has an equal and opposite pole:
\begin{equation}
    \Delta^{(p)} = \eval{ \Delta^{(p)}}_{\kappa_p \to \kappa_p^{\text{ren}}} -  \frac{1}{\epsilon}  \frac{(-)^k C_{m,n}^d}{k!}  \mu^{d+1 -2 (m+n)}\ .
\end{equation}
In this way, the total contribution to the conformal dimension with \(\mu\)-scaling \(d+2 -2(m+n)\) is regular as $\epsilon$ is sent to \(0\),
\begin{equation}
  \Delta^{(m,n)} + \Delta^{(p)} = Q^{(d+1-2(m+n))/d} \pqty{b_0 + b_1 \log(Q)} \ ,
\end{equation}
and only depends on the two Wilsonian parameters \(b_0\) and \(b_1\), 
which are part of the initial data of the \ac{eft}.
The cancellation of the poles thus leads to non-trivial tree-level $\log(\mu) \sim \log(Q)$ terms that have no counterpart in relativistic \acp{cft} at large charge.
Similar terms, however, do appear when one considers parity-violating CFT \cite{Cuomo:2021qws} or spinning operators \cite{Cuomo:2017vzg,Cuomo:2019ejv}.

\subsection{The universal one-loop logarithm in odd \(d\)}
\label{sec:universal-one-loop}

\noindent{}With the structure of the tree-level terms established, we can move on to the study of quantum corrections. The leading quantum contribution to the conformal dimension $\Delta(Q)$ enters at order $Q^0$ \cite{Hellerman:2020eff}. 
This comes from the Casimir energy of the fluctuations over the ground state, whose spectrum is given by~\cite{Kravec:2018qnu}
\begin{align}
  E^d_{n,l} &= \sqrt{\frac{4 n}{d} (n + l + d - 1) + l}\ ,  & n, l &= 0, 1, \dots ,
\end{align}
with multiplicities $M^{d-1}_l$ enumerated by the modes on the \(S^{d-1}\) sphere:
\begin{equation}
  M^{d-1}_l = \pqty{ 2l + d -2} \frac{\Gamma(l+d-2)}{\Gamma(l+1) \Gamma(d-1)}\ .
\end{equation}
The Casimir energy is then given by\footnote{The prime indicates that the zero mode $n = l = 0$ is excluded.}
\begin{equation}
  \Ecasimir{d} = \frac{1}{2} \sideset{}{'}\sum_{n,l=0}^{\infty} M^{d-1}_l E^d_{n,l}\ .
\end{equation}

This expression is clearly divergent and needs to be regulated.
Generally, there are two possibilities:  Either this is the \emph{only} contribution entering at order \(Q^0\), whereby the result of regularization must be finite and constitutes a \emph{universal} prediction of the theory, or there are other tree-level terms which appear with unknown Wilsonian coefficients.
In the case at hand (in odd $d$), the Casimir energy can be singular and be renormalized by tree-level contributions,
as with relativistic systems~\cite{Cuomo:2020rgt}.
When this happens, the renormalization results in the presence of a \emph{universal} $Q^0 \log(Q)$ term.

We have seen that tree-level terms contribute to the conformal dimension with \(\mu\)-scaling determined by Eqns.~\eqref{eq:mu-scaling-O-mn} and \eqref{eq:Zp_muscaling}. 
Edge operators never contribute at order $Q^0$, while bulk operators can only do so when $d$ is odd.
In particular, in \(d=2\), the regularization of the Casimir energy results in a universal prediction which was computed in~\cite{Hellerman:2020eff,Orlando:2020idm}:
\begin{equation}
\label{casimir2D}
    \Ecasimir{d=2} = -0.294159\dots
\end{equation}
In $d=3$, on the other hand, the three operators
\begin{equation}
\begin{aligned}
	\Op{2}{0} &= (\nabla U)^4 U^{-7/2}\ , &
	\Op{1}{1} &= (\nabla U)^2 Z U^{-5/2}\ ,\\
	\Op{0}{2} &= Z^2 U^{-3/2}\ ,
\end{aligned}
\end{equation}
contribute at order $Q^0$.
As we shall see shortly, they can be used to regularize the simple pole in the $\zeta$-regulated Casimir energy.

More generally, if the Casimir energy in $d+2\epsilon$ has a simple pole for \(\epsilon = 0\), \emph{i.e.,}
\begin{equation}
  \Ecasimir{d+2\epsilon} = \frac{E_{-1}}{\epsilon} + E_0 + \order{\epsilon}\ ,
\end{equation}
we can use bulk operators \(\Op{m}{n}\) satisfying \(m+n = (d+1)/2\) as counterterms in \ac{dimreg}.
We can write their cumulative leading-order contribution to $\Delta(Q)$ as
\begin{equation}
\begin{aligned}
  \Delta^{(m,n)} & = C_{m,n}^{d+2\epsilon} \mu^{2\epsilon} \\
  & = C_{m,n}^d \pqty{ 1 + \pqty{2 \log(\mu) + \text{cst.}} \epsilon + \order{\epsilon^2}}\ ,
\end{aligned}
\end{equation}
which we know to be regular as \(\epsilon \to 0\) from the analysis in the previous section.

The individual Wilsonian coefficients of these operators can be renormalized 
arbitrarily by
\begin{equation}
    c_{m,n} = c_{m,n}^{\text{ren}} + \frac{c_{m,n}^{(-1)}}{\epsilon}\ ,
\end{equation}
where $c_{m,n}^{\text{ren}}$ is regular, provided the following overall condition is satisfied:
\begin{equation}
  C_{m,n}^{d+2\epsilon} = \eval{C_{m,n}^{d+2\epsilon}}_{c_{m,n}\to c_{m,n}^{\text{ren}}} - \frac{E_{-1}}{\epsilon}\ .
\end{equation}
In concert with the structure of the classical divergences, the result is that the divergence in the Casimir energy is traded for a logarithm:
\begin{equation}
  \begin{aligned}
    \lim_{\epsilon \to 0}(\Ecasimir{d+2\epsilon} + \Delta^{(m,n)}) &=  \text{const.} - 2 E_{-1} \log(\mu) \\
    &= \text{const.} - \frac{2}{d} E_{-1} \log(Q)\ . 
    \label{eq:general-one-loop-log}
  \end{aligned}
\end{equation}
The novelty with respect to the discussion in the previous section is that \(E_{-1}\) can only come from the Casimir energy, is thus universal, and can be computed to yield a concrete prediction of the theory (as is the case in relativistic systems~\cite{Cuomo:2020rgt}).

The Casimir energy contains higher-order corrections in \(1/Q\), of course, which generically will have to be regulated and can give logarithms multiplied by negative powers of $Q$ in the conformal dimension.
In any case, they will have to be regulated by bulk or edge operators that are not divergent (only bulk operators with positive \(\mu\)-scaling can be divergent), so we do not face situations in which a divergence in the Casimir energy is to be regulated by a bulk operator that must, itself, be regulated by an edge term.
This excludes the presence of double logarithms coming from one-loop effects in the expansion.

\subsection{The universal one-loop logarithm in $d=3$}

\noindent{}In this section, we compute the universal \(Q^0 \log(Q)\) contribution to the conformal dimensions in the special case \(d = 3 + 2 \epsilon\) using  \ac{dimreg}.

\begin{widetext}
The Casimir energy is
\begin{equation}
  \Ecasimir{3+2\epsilon} = \frac{1}{2} \sideset{}{'}\sum_{n,l=0}^{\infty} M^{2+2\epsilon}_l E^{3+2\epsilon}_{n,l}\ .
\end{equation}
In the limit \(\epsilon \to 0\), the multiplicity is
\begin{equation}
  M_l^{2 + 2\epsilon} = \pqty{2l + 1} + 2\left[ \pqty{2l +1} H_l - 2l \right] \epsilon + \order{\epsilon^2}\ ,
\end{equation}
where \(H_l\) is the \(l\)-th harmonic number.
We are interested in the behavior around the pole, which is completely determined by the large-\(l\) behavior of the sum.
We can thus further expand
  \begin{equation}
      M_l^{2 + 2\epsilon} = \pqty{2 l + 1} + 2\pqty{ \pqty{2l+1} \pqty{\log(l) + \gamma - 1} + \order{\frac{1}{l}} } \epsilon %
      = \pqty{2 l + 1 + 4 \epsilon} \pqty{l^{2\epsilon} + \pqty{\gamma -1} 2\epsilon} + \order{\frac{\epsilon}{l} }~.
  \end{equation}
  Up to terms that vanish in the \(\epsilon \to 0\) limit, the Casimir energy sum is given by

    \begin{equation}
    \Ecasimir{3+ 2\epsilon} = \frac{1}{2} \sideset{}{'}\sum_{n,l=0}^{\infty} \pqty{2l + 1} l^{2\epsilon} \pqty{ \pqty{n + \frac{3}{4} } \pqty{l + n +\frac{5}{4} } - \frac{15}{16} }^{1/2} + \text{regular for \(\epsilon \to 0\)~.}
  \end{equation}

  This expression readily yields to zeta-function regularization.
  Applying the binomial expansion four times in a row, we can rewrite the sum in terms of the Mordell--Tornheim zeta function $\zeta_{MT}(s_1, s_2, s_3)$, defined as~\cite{mordell1958evaluation,tornheim1950harmonic,matsumoto2003mordell}
  \begin{equation}
    \zeta_{MT}(s_1, s_2, s_3) \equiv \sum_{n,l=1}^\infty l^{-s_1} n^{-s_2} (n + l)^{-s_3}~.
  \end{equation}
  We obtain:
  \begin{multline}
    \Ecasimir{3+ 2\epsilon} = \left( \frac{3}{4} \right)^{-\sfrac{1}{2}} \sum_{k,k_1,k_2,k_3=0}^\infty \binom{\sfrac{1}{2}}{k} \binom{1}{k_1} \binom{\sfrac{1}{2}-k}{k_2} \binom{\sfrac{1}{2}-k}{k_3} 
	 \left( -\frac{15}{16} \right)^k \left( -\frac{1}{2} \right)^{k_1} \left( -\frac{1}{4} \right)^{k_2} \left( -\frac{3}{4} \right)^{k_3} \\
    \times \zeta_{MT}(k_1 -2 \epsilon - 1 ,  k + k_2 - \sfrac{1}{2} , k + k_3 - \sfrac{1}{2}) + \text{regular.}
  \end{multline}
  The analytic structure of this special function is known~\cite{matsumoto2002analytic,tsumura2005mordell,matsumoto_value-relations_2008}, and from the expansion
  \begin{multline}
    \zeta_{MT}(s_1, s_2, s_3) = \frac{\Gamma(s_2 + s_3 -1) \Gamma(1 - s_2)}{\Gamma(s_3)} \zeta(s_1 + s_2 + s_3 -1) 
    + \sum_{k=0}^{M-1} \binom{-s_3}{k} \zeta(s_1 + s_3 + k) \zeta(s_2 - k) + \text{regular,}
  \end{multline}
\end{widetext}
we see that there are poles with unit residue if\footnote{Note that when $s_2$ is a positive (non-zero) integer, the expression admits two singular pieces that cancel each other.}
\begin{equation}
  \begin{cases}
    s_1 + s_3 = 1, 0, -1, -2, \dots \\
    s_2 + s_3 =  1, 0, -1, -2, \dots \\
    s_1 + s_2 + s_3 = 2~.
  \end{cases}
\end{equation}
In our case, only the third channel in this set can produce singularities, which arise for
\begin{equation}
  k_1 + k_2 + k_3 + 2 k -2 \epsilon = 4~.  
\end{equation}
Summing all the contributions, we find the final result
\begin{equation}
  \Ecasimir{3 +2 \epsilon} = -\frac{1}{ 2\sqrt{3} \epsilon} + \text{regular.}
\end{equation}
Using the general formula in Eq.~\eqref{eq:general-one-loop-log}, we thus find that the conformal dimensions have a universal \(Q^0 \log(Q)\) term given by
\begin{equation}
\label{Q0LogQin3d}
  \eval{\Delta(Q)}_{Q^0}  = \frac{1}{3\sqrt{3}} \log(Q) + \text{const.}
\end{equation}

\section{Connected multi-vertex contributions and corrections to the classical solution}
\label{sec:multi-vertex}

\noindent{}Thus far, we have analyzed the classical and one-loop vacuum-energy contributions
of single insertions of operators.  We have not addressed
the connected terms coming from multiple insertions.  At one loop, all connected terms with more than one insertion are necessarily non-1PI.  For a translationally invariant background of a translationally-invariant theory, non-1PI vacuum diagrams are
necessarily trivial, removable by a renormalization of the one-point vertex at zero momentum.  For an inhomogeneous background, such as the one
we are considering, however, corrections to the one-point function from tree- and loop-level contributions are generally nontrivial.  As a result, non-1PI diagrams are necessarily nontrivial as well.

There are two important aspects to consider.  First, we must determine whether non-1PI terms spoil the tree-level $\mu$-counting, which we have inferred from an analysis of the classical contributions of single insertions.  Second, we should consider
the issue of tree-level non-1PI vacuum corrections, together with the question of higher-derivative contributions to the classical profile of
the charge distribution.

For the first of these questions, we can appeal to the analysis in~\cite{Hellerman:2020eff}.
There, the nonlinear
contribution of operators without leading-order classical expectation values is considered.  To obtain a bound on the $\mu$-scaling exponent of such 
terms, the authors consider an unspecified cutoff
for the \ac{eft}, such that the theory is still weakly
coupled at the cutoff, even when the parametrically
lower local \ac{uv} scale near the boundary is taken into
account.  By breaking up the field into \ac{vev}s and fluctuations, one can consistently assign an upper limit
to the $\mu$-scaling exponent of a quantum fluctuation
of the $\chi$-field.  In this way, one bounds the $\mu$-scaling of an operator nonlinearly at the classical or quantum level, even when the nonlinear or quantum
contribution of the operator is its \emph{leading} effect
on the vacuum energy. 

Concretely: In general $d$ dimensions for $n_{i}$ insertions of a dressed and integrated \(Z\) operator, the upper bound for the total $\mu$ scaling is $d + 1 - n_i(d/3 + 2)$. \emph{E.g.,} in $d=2$, we get $3 - 8/3 n_i$ as in~\cite{Hellerman:2020eff}.     
Until $d = 9$, the leading contribution (with $n_i=2$) always enters with
negative $\mu$ scaling. 
The upper limit on the $\mu$-scaling derived this way
is rather coarse, passing over several more refined
considerations that would likely improve the bound.  Nonetheless, the bound is strong enough to rule out
the possibility of such operators contributing to the vacuum energy with non-negative $\mu$-scaling in $d\leq 9$ spatial dimensions, proving the universality of
the one-loop correction in those dimensions, including the $Q^0$ term in $d=2$ and the $Q^0 \log(Q)$ term $d=3$ (Eqns.~\eqref{casimir2D} and \eqref{Q0LogQin3d}). 

Note that we have not analyzed whether multiple-insertion, non-1PI classical ground-state contributions could contain logarithms of $\mu$ with negative $\mu$-scaling.
Of course, there already exist sources of $\log(Q)$ contributions with negative $Q$-scaling from classical effects, coming from the logarithmically-enhanced subleading power-law terms in the relationship between $Q$ and $\mu$.  These must exist due to the presence of $\log(\mu)$-enhanced subleading terms in the classical formula for the ground state energy as a function of $\mu$.  
In short, the question of which negative $Q$-scaling terms might give rise to logarithmic enhancements, and, for that matter, the broader study of two- and higher-loop quantum corrections, are issues left for future investigation. 

Finally, another aspect of the vacuum non-1PI contributions worth noting is that they are directly related to classical and quantum corrections to the one-point function of $\chi$.  For purposes of computing subleading large-$Q$ corrections to
the ground state energy, it is not necessary to compute these corrections to the $\chi$ profile explicitly; they are implicitly
incorporated by the inclusion of non-1PI diagrams.  Nonetheless, it would
be interesting to compute directly the
quantum and higher-derivative corrections to the $\chi$ profile or, equivalently, the
density profile, as the density profile
has been measured experimentally \cite{PhysRevLett.92.120401,Giorgini_2008}.%
\footnote{Unlike in the relativistic case, the \(\chi\) profile depends on subleading terms with Wilsonian coefficients generalizing Eq.~\eqref{eq:LO-action}, and it is not completely fixed by the symmetry of the problem.}

\section{Conclusions}

\noindent{}In this note, we have analyzed the ground-state energy of a \ac{nrcft} in the Son--Wingate large-charge universality class, working in the large-charge expansion.
Specifically, we have elucidated the conditions for the appearance of logarithmic divergences in the expansion, for classical and one-loop contributions to the energy.  Dimensional regularization provides a conformally invariant renormalization scheme that renders the calculation tractable.  Renormalization transmutes the logarithmic divergences into \(\log(Q)\) enhancements of power-law terms in the expansion of the energy and operator dimension.  

In $d$ spatial dimensions, we have demonstrated distinct patterns in the appearance of \(\log(Q)\) terms that depend on whether the spatial dimension $d$ is odd or even. 
In odd dimension $d$, we find that logarithmic divergences from \emph{classical} contributions to the ground state energy are absent, along with their associated \(\log(Q)\) terms in the energy. In even spatial dimension $d$, we find that such contributions can appear, associated with divergent integrals of classical terms near the edge of the droplet charge distribution.  These classical \(\log(Q)\) terms are always multiplied by positive powers of $Q$. 
In any even spatial dimension $d$, the particular positive powers of $Q$ multiplying the \(\log(Q)\) term are restricted to a limited finite set, defined by the cases where a bulk local operator and an edge operator share the same $Q$-scaling.  We have provided a table of the particular $Q$-scalings where such logarithmic enhancements can occur in various even spatial dimensions (Table~\ref{tab:divergent-bulk-operators}); this pattern of classical \(\log(Q)\) enhancements is illustrated in Figure~\ref{fig:Lines-and-dots}.  

We have also analyzed the appearance of \(\log(Q)\) terms in the vacuum energy coming from one-loop Casimir contributions to the ground state energy.  Using dimensional regularization to make the analysis tractable, we have found that there are never
any $Q^0 \log(Q)$ terms in even spatial dimension $d$.  In odd spatial dimension $d$ there is a $Q^0 \log(Q)$ term with a calculable coefficient.  This is precisely
analogous to the situation for the large-charge expansion in $(d+1)$-dimensional relativistic \ac{cft}, as first observed 
by~\cite{Cuomo:2020rgt}.  At sub-leading order, $\log(Q)$ terms multiplied by negative powers of $Q$ can also
arise from contributions of higher-derivative terms in the action to the one-loop diagram.
Importantly, in $d=3$, 
we have provided a concrete experimental prediction for the quantum behavior of the energy spectrum of a harmonically trapped unitary Fermi gas with large particle number.  In particular, we have proven the existence of a $\frac{\omega}{3 \sqrt{3}} Q^0 \log(Q)$ term in the ground-state energy.

The results presented here draw upon the rich structure inherent to conformal field theories, while relying on the dramatic simplifications afforded by large-quantum-number perturbation theory. Of course, much of this technology traces its parentage to a number of endeavors emerging from string theory.\footnote{{\it E.g.,\rm} see \cite{Berenstein:2002jq, Alday:2007mf, Komargodski:2012ek, Fitzpatrick:2012yx, Alday:2016njk, Giombi:2021zfb}.} As emphasized above, if cold-atom systems in the unitary regime can be studied with sufficient precision, they may provide a crucial terrestrial test-bed for this increasingly far-reaching arena of technology in theoretical physics.

\begin{widetext}
  \section*{Acknowledgments}

  \noindent{}S.H. thanks the Simons Center for Geometry and Physics and the Aspen Center for Physics for hospitality while
  this work was in progress.  The work of S.H. is supported by
  the World Premier International Research Center Initiative (\textsc{wpi} Initiative), \textsc{mext}, Japan;
  by the \textsc{jsps} Program for Advancing Strategic International Networks to Accelerate the
  Circulation of Talented Researchers; and also supported in part by \textsc{jsps} \textsc{kakenhi} Grant
  Numbers \textsc{jp}22740153, \textsc{jp}26400242.
  The work of V.P. and S.R. is supported by the Swiss National Science Foundation under grant number 200021 192137.
  D.O. is partly supported by the \textsc{nccr 51nf40--141869} ``The Mathematics of Physics'' (Swiss\textsc{map}).
  We also thank Shauna Kravec and Nathan Lundblad for illuminating discussions on cold-atom systems.
\end{widetext}

\bibliographystyle{utphys}
\bibliography{References,not-arxiv}

\end{document}